\documentclass[12pt,preprint,letterpaper]{aastex}
\usepackage{graphicx}

\begin{document}

\title{The UV--mid-IR Spectral Energy Distribution of a $z=1.7$ Quasar Host Galaxy}
\author{N.R.~Ross\altaffilmark{1},
R.J.~Assef\altaffilmark{1},
C.S.~Kochanek\altaffilmark{1,2},
E.~Falco\altaffilmark{3},
S.D.~Poindexter\altaffilmark{1}
}

\affil{
  \altaffiltext{1} {Department of Astronomy, The Ohio State 
  University, 140 W.\ 18th Ave., Columbus, OH 43210}
  \altaffiltext{2} {Center for Cosmology and Astroparticle Physics,
  The Ohio State University, 191 W.\  Woodruff Avenue, Columbus, OH 43210}
  \altaffiltext{3} {Harvard-Smithsonian Center for Astrophysics, 60 Garden Street, Cambridge, MA 02138}
}

\begin{abstract}
We have measured the spectral energy distribution (SED) of the host galaxy
of the $z_s=1.7$ gravitationally lensed quasar SDSS~J1004+4112 from
$0.44-8.0\mu$m ($0.16-3.0\mu$m in the rest frame).  The large angular
extent of the lensed images and their separation from the central
galaxy of this cluster lens allows the images to be resolved even
with the Spitzer Space Telescope.  Based on the SED, the host
galaxy is a mixture of relatively old and intermediate age stars
with an inferred stellar mass of $\log (M_{\star}/M_{\sun})=11.09\pm0.28$ 
and a star formation rate of $\log(\dot{M}/$M$_{\odot}$~yr$^{-1})=1.21\pm0.26$.
Given the estimated black hole mass of $M_{BH}\simeq
10^{8.6} $M$_{\odot}$ from locally-calibrated correlations of
black hole masses with line widths and luminosities, the black hole
represents a fraction $\log(M_{BH}/M_{\star}) = -2.49\pm0.28$ of the stellar mass and it
is radiating at $0.24\pm0.05$ of the Eddington limit.  The ratio of
the host stellar mass to the black hole mass is only marginally consistent with the locally observed ratio.
\end{abstract}


\section{Introduction}\label{sec:intro}

In the local universe, the host galaxies of active, luminous black holes
tend to be bluer star forming galaxies with a roughly 1000:1 ratio between
star formation and accretion rates \citep{kauffmann05}.  More luminous
AGN also show younger stellar populations.  Moreover, the relative growth
rates match the observed local ratio of stellar to black hole mass
\citep{kauffmann05,marconi03}.  At higher redshifts, $z > 1$, the picture is less clear because
the greater distances and higher typical AGN luminosities make it increasingly
difficult to study host galaxies. Studies by \citet{peng06a,peng06b} 
argue that the relationship is shifted and that at this epoch ($z>1$) the
black hole mass grows faster relative to the stellar mass than is
observed locally, while \citet{lauer07} and \citet{dimatteo08} argue for little change.

There is also considerable interest in the star formation rates of the
hosts at these redshifts.  With the now prevalent view that the black
holes and stars grow in a self-regulating process \citep[e.g.][]{hopkins05a,hopkins05,hopkins06b,sijacki07,dimatteo08}, particularly
during major mergers, it is of considerable importance to be able to
estimate both the stellar mass and the star formation rate.  In the
\citet{hopkins05a,hopkins05} scenario, the peak star formation rates precede
the peak quasar luminosity, and the quasar phase lasts about $10^7$
years.  Unfortunately, estimating both stellar masses and
star formation rates at these redshifts requires not only a detection of the host but
a reasonably complete spectral energy distribution (SED).

Here we make use of gravitational lensing to measure the SED of a
$z_s=1.73$ quasar host galaxy from $0.44$--$8.0\mu$m and infer its mass and
star formation rate.  As emphasized by \citet{peng06b}, 
quasar lenses are ideal laboratories for studies
of quasar hosts because the lens magnification ``pulls'' the
host out from under the quasar to provide a $\sim10^2$ improvement
in contrast.  Moreover, the arced shapes of the lensed hosts
are easily distinguished from PSF artifacts.  Our target is
the five image lens SDSS~J1004+4112 \citep{inada03,inada05,sharon05,ota06,fohlmeister07,fohlmeister08,inada08}.  This lens is
created by a $z_l=0.68$ cluster of galaxies \citep{inada03}, giving it 
exceptionally large image separations (a $\sim14$~arcsec Einstein
ring diameter) that both lead to very large images of the host and
places the quasar images well away from the lens galaxy emission.
In fact, the host is so extended and well-separated from
the lens that it can be resolved by the Spitzer Space
Telescope with relative ease.
There are also additional, higher redshift lensed galaxies
\citep{sharon05} and time delay measurements between the various lensed images of the quasar of $\Delta \tau_{BA}=40.6\pm1.8$ days,
$\Delta \tau_{CA}=821.6\pm2.1$ days, and $\Delta \tau_{AD}>1250$ days \citep{fohlmeister08}.
In \S \ref{sec:data} we describe how we measure the 
spectral energy distribution (SED) of the host galaxy and the 
quasar. In \S \ref{sec:analysis} we use these estimates to 
determine the luminosity, stellar mass, and star formation 
rate of the host galaxy. We assume a flat, $\Omega_0=0.3$, $H_0=70$~km/s/Mpc cosmology.

\section{Data}\label{sec:data}

We use Hubble Space Telescope (HST) and Spitzer Space Telescope (SST) observations of 
SDSS~J1004+4112 in 8 bands covering the visual to mid-infrared wavelengths. The HST 
data consists of ACS/WFC B(F435W), V(F555W), I(F814W) observations and NICMOS/NIC2 
H(F160W) observations. The SST/IRAC data consists of 3.6, 4.5, 5.8, and 8.0$\mu$m 
observations.  For the V, I, H, and IRAC bands we have multiple observational epochs.  
A summary of the observations is given in Table \ref{tab:obs}. Each HST observation
consists of several (typically 4) sub-exposures, drizzled together \citep{fruchter02} to
create one background-subtracted image.  Each IRAC epoch consists of 36 dithered 96.8 second 
images in each of the IRAC channels.  Starting with the Basic Calibrated Data frames, the mosaic is oversampled by a factor of 4 using  
MOPEX (Mosaicking and Point Source Extraction), with outlier rejection 
to remove cosmic rays.  Typical image depths (in Vega magnitudes) are given in Table 
\ref{tab:obs}.

We first used a parametric model to fit the images using a combination of point sources 
for the quasars, exponential disks and de Vaucouleurs models for the cluster galaxies and 
Gaussians for the images of the host galaxies, all convolved with point spread function (PSF) models, as in \citet{lehar00}. 
These have problems for estimating the flux of the host galaxy due to the fact that the PSF, 
generated by TinyTim for the HST bands and obtained from SST for the IRAC bands, has 
significant fractional errors at the peak of the quasar, exactly where the model for the 
host galaxy also peaks.  The parametric models tend to overestimate the flux of the host 
galaxy in order to reduce the residuals at the position of the quasar. We will use these 
models only to correct aperture magnitudes for the effects of the point spread 
function.

\begin{figure}
  \begin{center}
    \plotone{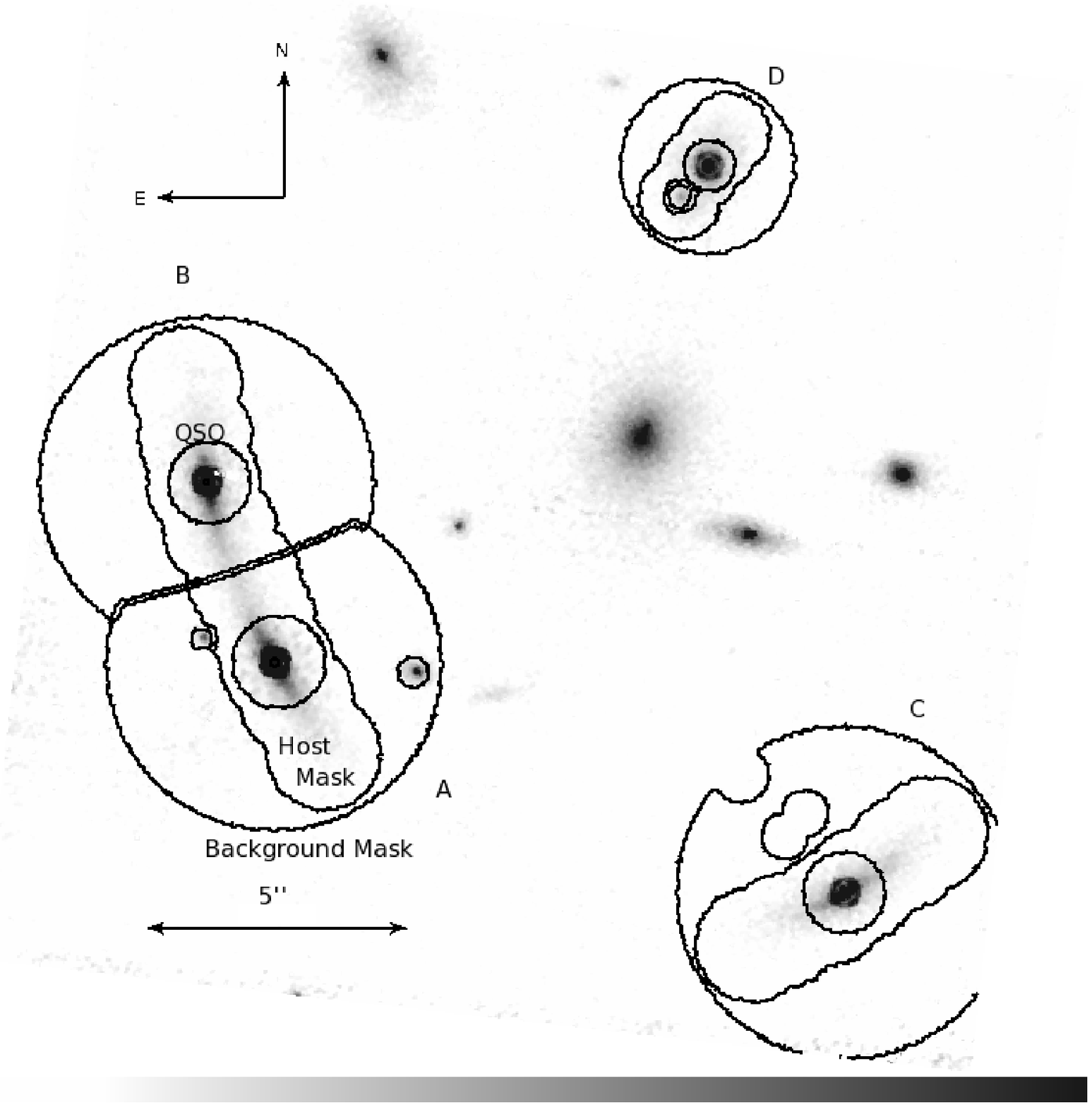}
    \caption{NICMOS/NIC2 H-band (F160W) image of SDSS J1004+4112 with mask outlines denoted 
by the solid black lines. The host galaxy is clearly seen stretched out from beneath the 
peak of the QSO. For image D we only use the joint mask.}
    \label{fg:Hmasks}
  \end{center}
\end{figure}

We next created a series of masks which isolate regions on the images where the flux is 
dominated by either the host galaxy or the quasar, in both cases excluding flux from objects 
in the field. These masks have regions with value either 0 or 1 in order to exclude or 
include flux, respectively, in specific pixels when multiplied into the original images. We keep the masks consistent across all 
bands by geometrically transforming a master copy to the appropriate centering, pixel scale, 
and orientation of each observation. We defined three types of masks. Host masks exclude 
flux both near the quasars and away from the host images seen in the I/H data. Quasar masks 
include only flux near the peak of the quasar images. Background masks include a region 
outside the host mask which we use to estimate any residual background flux. Joint masks 
combine the host and the quasar masks to estimate the total flux of both components. 
Figure \ref{fg:Hmasks} superposes these masks on an H-band image.

When we apply a mask to a region, we calculate the flux $f_{mask}$ under the mask. This flux 
is a combination of the true flux in the masked region $f$ with contamination $f_{cont}$ 
spread into the masked region by the PSF, losses $f_{loss}$ out of the region due to the 
PSF, and $f_{back}$ due to any mis-estimation of the background during the image reduction 
process. For example, we estimate $f_{cont}$ and $f_{loss}$ for our host mask as follows. 
We start from the model of the image without PSF convolution. We then mask this image, 
convolve it with the PSF, and measure the flux under the second mask. Thus the  
contamination, $f_{cont}$, of the host mask region due to the PSF spreading flux out of the quasar mask 
region, is found by first masking the unconvolved model image with the quasar mask, 
convolving this masked image with the PSF, and then measuring the flux found in the host mask region, while $f_{loss}$ is found by 
masking with the host mask, convolving with the PSF and then measuring the flux outside of 
the host mask.  Since these corrections are modest, we are not very sensitive to the problems 
in the model image. We estimate the background by subtracting the model from the data and 
measuring the residuals in the background mask. The resulting flux for any region is then
\begin{equation}\label{eq:flux}
f\ =\ f_{mask}\ -\ f_{cont} \ +\ f_{loss}\ -\ f_{back}
\end{equation}
The measurements are summarized in Table \ref{tab:fluxtab}. We first estimate statistical 
uncertainties in the magnitudes using a bootstrap resampling of the images. The bootstrap 
resampling technique creates an ensemble of trial images by sampling with replacement from 
the sub-images (dithers, CR Splits, etc.) that were averaged together for each observation.
We analyze each trial image in the same manner as the true images and estimate error bars from 
the variance of the results over the trials.
The remaining uncertainty arises from the background subtraction.
We recompute all the estimates using two different regions for
estimating the background flux, as well as a background estimate 
generated by the model fits.  The dispersion of these background 
estimates, multiplied by the number of pixels in a given mask 
region, gives an estimate of the background uncertainties in each mask region. Small changes in the estimated
background can have significant effects on the flux measured for the host because of
the large number of pixels in both the host mask and the joint mask.
The uncertainties we present in Table 
\ref{tab:fluxtab} are a combination of these statistical and background uncertainties, added in quadrature.

We use data from the ongoing monitoring of SDSS~J1004+4112 \citep{fohlmeister08} to correct for 
time variability of the quasar in the QSO and joint masks. We chose 13 December 2005 as the reference 
date, as many of our observations were made close to this date (see 
Table~\ref{tab:obs}).  We estimate the time delay corrections by comparing the flux measured in 
the monitoring project on or within $\sim$2 days of the observation with the flux measured in the 
monitoring project on or within $\sim$2 days of the reference date.  These time delay corrections range 
from about 0.05 to $-0.44$ magnitudes.  We use these time delay corrections only for the optical to 
near-IR observations of the quasar because the observations in the mid-IR show much less variability, 
as one would expect from the general trend of reduced variability at longer wavelengths 
\citep[e.g.][]{vandenberk04}. We do not correct for the time delays between the lensed images because 
the image D time delay is not known \citep[see][]{fohlmeister08}.  Essentially, we will ``time average'' 
the properties of the quasar in our final results.

\section{Analysis}\label{sec:analysis}

We used an extended version of the SED template models presented by \citet{assef08} to fit the data for 
each lensed image. These templates consist of early-type, Sbc, Irr and QSO templates empirically derived 
by fitting the GALEX UV through SST/MIPS 24$\mu$m SEDs of 13623 ``pure'' galaxies (with no obvious signatures 
of nuclear activity) and 4242 quasars and galaxies with AGN activity in the NDWFS Bo\"otes field \citep{januzzi99}, with redshifts 
measured by the AGN and Galaxy Evolution Survey \citep[AGES,][]{kochanek08}. \citet{assef08} details the 
procedure used to derive these templates, and they will be discussed in depth in an upcoming paper by 
\citet{assef09}.  A fit is produced by multiplying each template by a coefficient and finding the $\chi^2$ 
minimizing fit to the data over these coefficients, with the added constraint that all template coefficients 
must be non-negative, as subtracting a template is unphysical \citep[see the discussion in][]{assef08,assef09}.

\begin{figure}
  \begin{center}
    \plotone{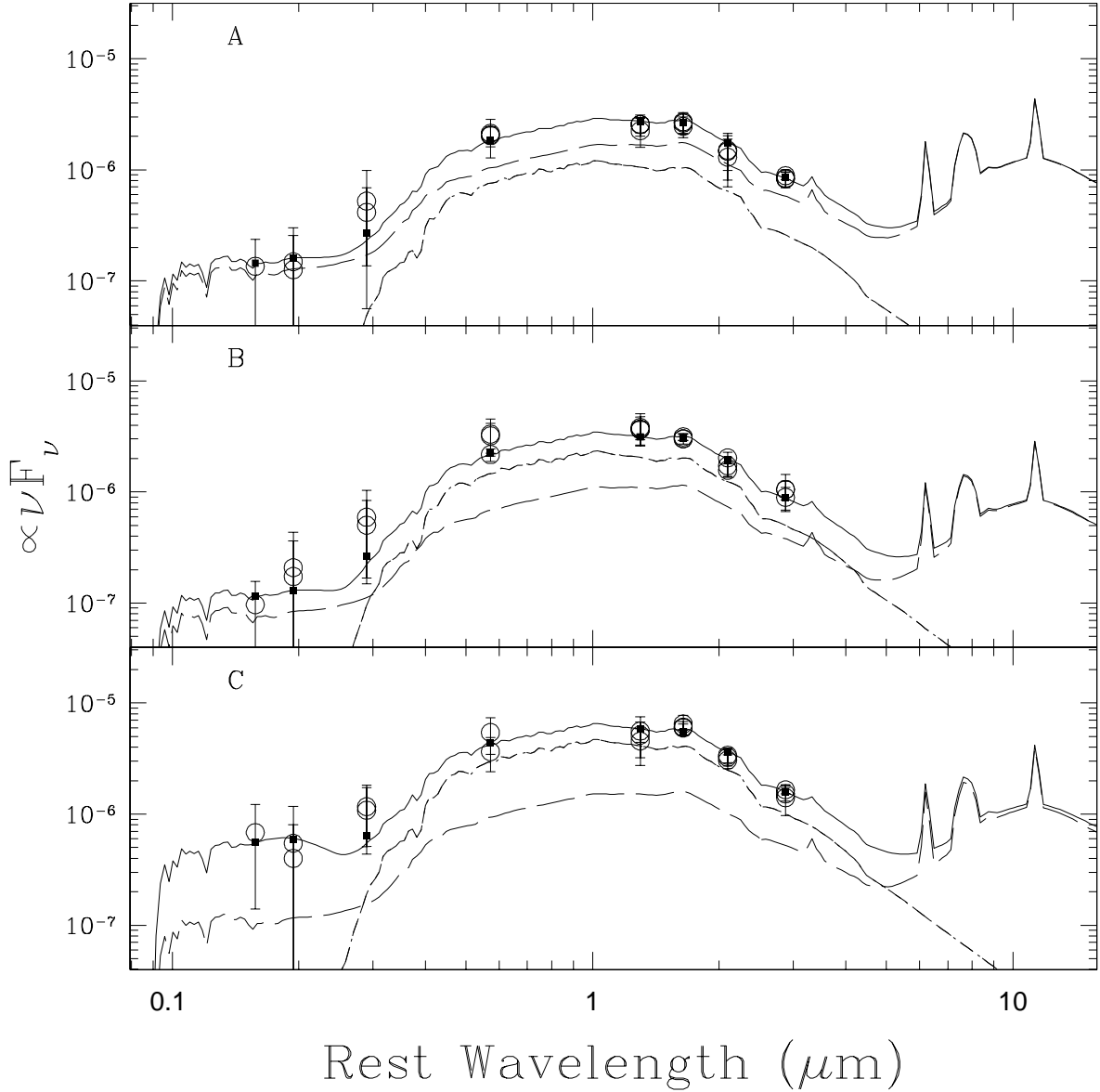}
    \caption{Host mask spectral energy distribution for images A (Top), B (Middle), and C 
(Bottom). Observation bands are, from left to right, ACS/WFC B(F435W), V(F555W), I(F814W), 
NICMOS/NIC2 H(F160W), IRAC 3.6, 4.5, 5.8, and 8.0 $\mu$m. The results for all observation 
epochs are shown. The solid, dot-dashed, long-dashed, and short-dashed lines correspond to 
the total SED and the contribution from the E, Sbc, and Irr templates respectively. The open 
circles are the measured fluxes while the closed squares are the best fit value given the template fits. The contribution from the Irr 
template is too small to be seen on this scale. The SEDs are corrected for magnification by the lens based on the models of \citet[][Oguri private communication]{inada08}.}
    \label{fg:seds}
   \end{center}
\end{figure}

\begin{figure}
  \begin{center}
    \plotone{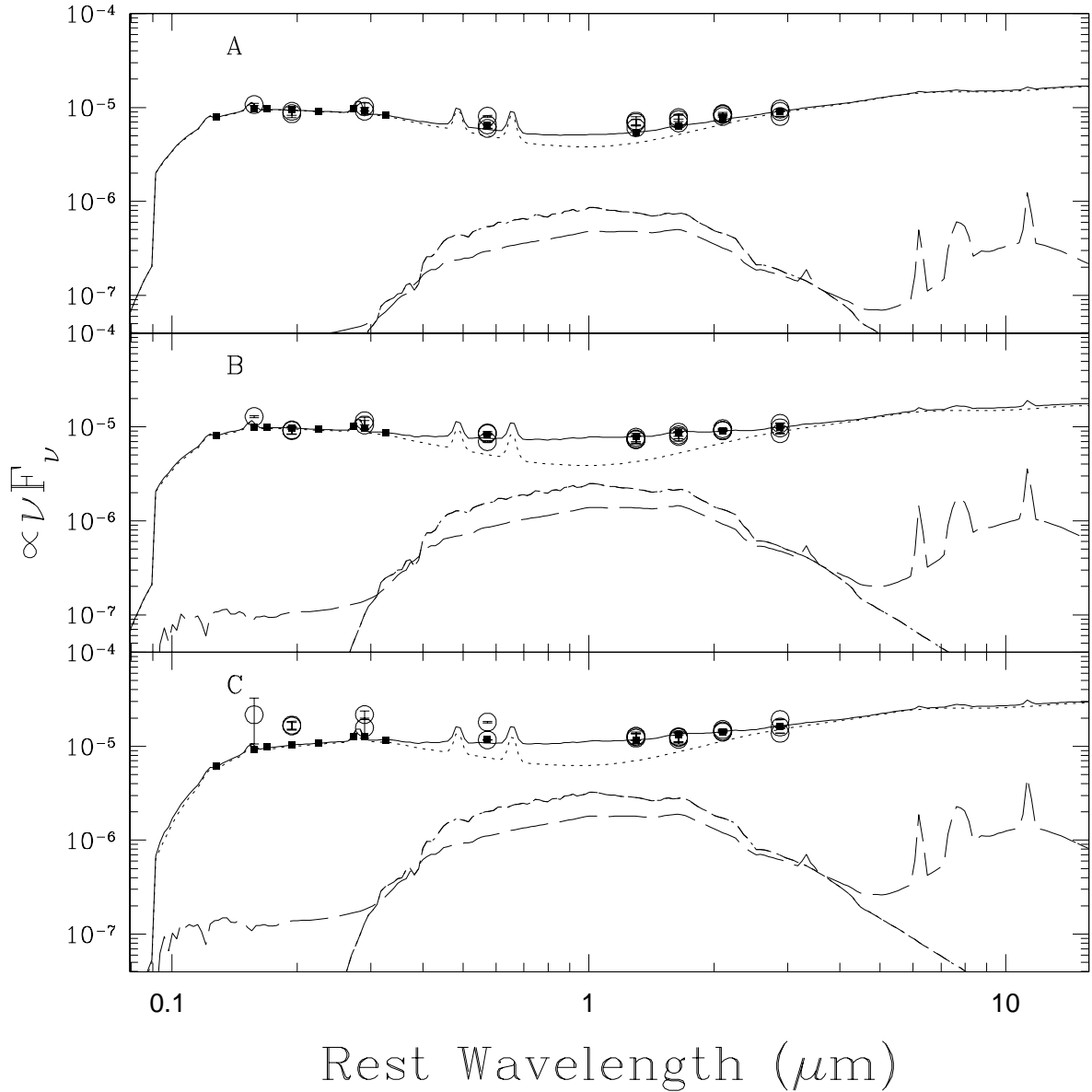}
    \caption{QSO mask spectral energy distribution for images A (Top), B (Middle), and C 
(Bottom). The galaxy templates and data points are represented as described in Figure 
\ref{fg:seds} while the dotted line shows the QSO template and the solid line shows the 
sum of all templates. The contribution from the Irr template is too small to be seen on this scale.
The changes in the optical continuum slope are due to the variations in the QSO extinction estimates.}
    \label{fg:sed_qso}
   \end{center}
\end{figure}

\begin{figure}
  \begin{center}
    \plotone{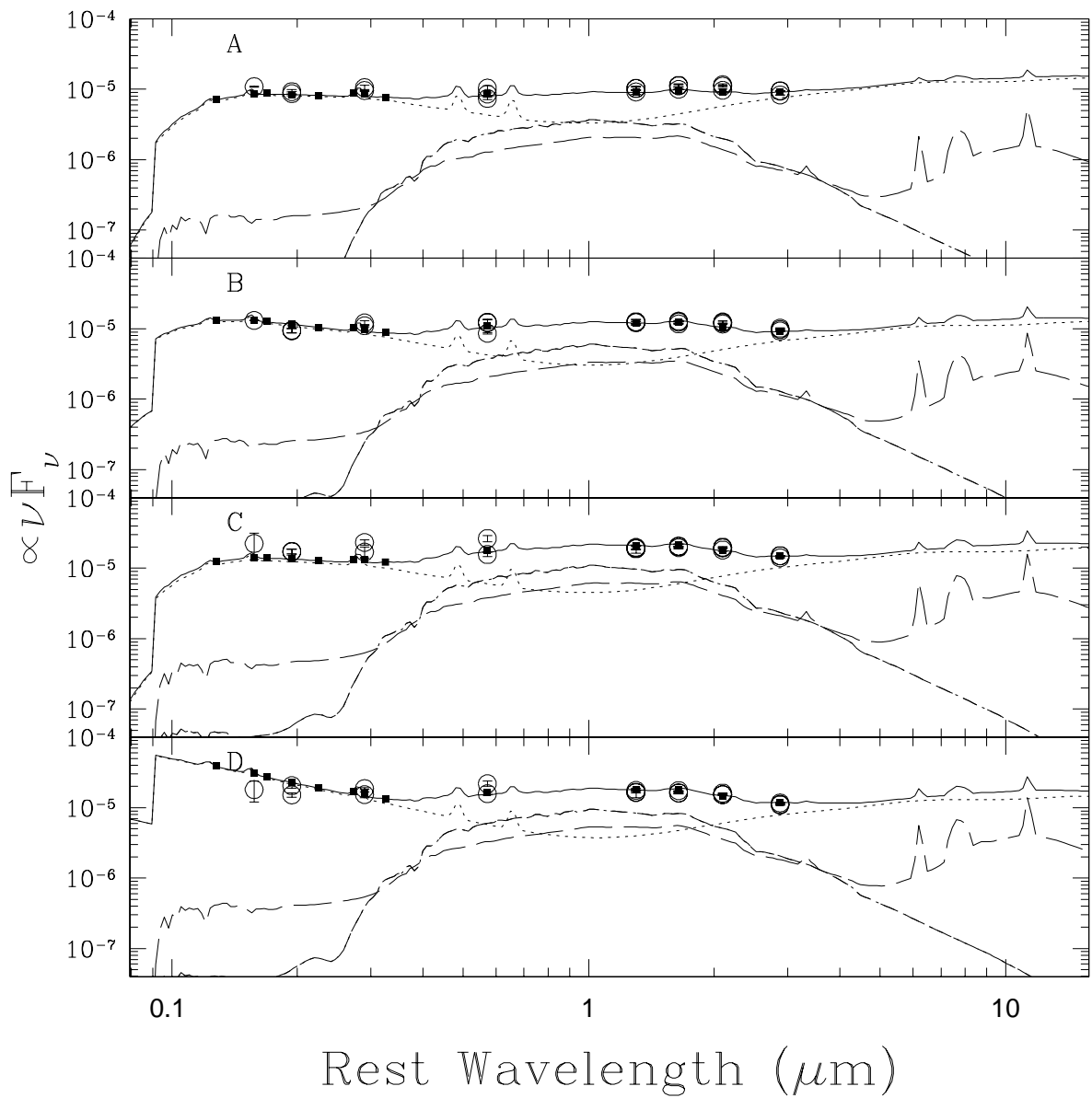}
    \caption{Joint mask (QSO + Host) spectral energy distribution for images A (Top), B (Top Middle), C 
(Bottom Middle), and D (Bottom). The components are as described in Figs. \ref{fg:seds} and 
\ref{fg:sed_qso}.   The contribution from the Irr template is too small to be seen on this scale. Note that the template models show 
some variation in the amount of extinction accounted for in the various images.}
    \label{fg:sed_joint}
   \end{center}
\end{figure}

For images A, B, and C we separately fit the host, QSO and joint SEDs, while for image D we only fit the 
joint SED.  Figures \ref{fg:seds}, \ref{fg:sed_qso}, and \ref{fg:sed_joint} show examples of the template 
fits and Table~\ref{tab:spectab} lists the host luminosities derived from the template fits to the host, 
QSO, and joint mask data. These luminosities are corrected for magnification by the lens using magnifications 
of 28.5, 19.1, 9.8, 7.8 for the A, B, C, and D images, respectively, from the models of Inada et al. (2008; 
Oguri, private communication).

We used several methods to estimate our systematic uncertainties in determining the properties of the host galaxy.  First, we fit the host 
mask data both with and without the B-, V-, and I-band data in order to examine whether eliminating the data points with the worst signal-to-noise 
ratios would produce any difference in the fits. These fits produced 
significant variations on an image by image basis, but showed little effect on the averages, which is unsurprising, considering that these data points are already heavily downweighted by their large uncertainties during the fitting process.  
Next we fit the QSO and joint mask data once by allowing all template components to vary 
and once by fixing the ratios of the galaxy templates using the results of the host mask 
fits. These fits produced different results for the template ratios (although the Irr template 
was never favored) and the joint fits have systematically brighter host and fainter AGN 
components than the simple sum of the separate host and quasar mask results, but these 
variations had no significant effect on the luminosities, masses, or star formation rates.

We use the template fits and standard scaling relations to estimate the stellar mass and star formation 
rate of the host galaxy.  We use the estimated rest frame 8.0$\mu$m flux of the host to estimate the total 
infrared luminosity based on the scalings of \citet{bavouzet08}, which in turn is used to estimate the obscured star 
formation rate (SFR)
\begin{equation}\label{eq:sfr}
\frac{SFR}{1 \mathrm{M}_{\Sun}\,\mathrm{yr}^{-1}}=\frac{L_{FIR}}{5.8\times10^9 \mathrm{L}_{\Sun}}
\end{equation}
of the host galaxy using the local scalings of \citet{kennicutt98}, corrected for the difference 
in the definition of the total infrared luminosity between \citet{bavouzet08} and \citet{kennicutt98}. The 
uncertainties in the star formation rates are dominated by the uncertainties in extrapolating to 
the total infrared luminosity from the 8.0$\mu$m flux.  \citet{bavouzet08} found a ~38\% scatter 
between the 8.0$\mu$m flux and $L_{FIR}$, while  \citet{kennicutt98} found a scatter of about ~30-50\% 
between the $L_{FIR}$ and the SFR, and also attributed it to the uncertainty in estimating the 
FIR luminosity from the near infrared luminosity and uncertainty in the effects of extinction\citep[for further discussion of these uncertainties, see \S 6 of][]{caputi07}.  We also used 
the scaling relation in \citet{caputi07}, who argue for a different high-luminosity L$_{8.0\mu m}$-to-L$_{IR}$ 
relationship than \citet{bavouzet08}.  We found no significant difference between 
these two scaling relations, as the host galaxy has an 8.0$\mu$m luminosity close to the break luminosity 
where the \citet{caputi07} relation begins to differ from that of \citet{bavouzet08}.
We then used the \citet{kennicutt98a} relation between UV luminosity and SFR to produce an estimate of the unobscured SFR.  The \citet{kennicutt98a} relation converts the UV luminosity--SFR relation of \citet{madau98} to a Salpeter IMF with continuous star formation.  Note that the \citet{kennicutt98a} relation would not apply for a starburst galaxy.  The Salpeter IMF used in \citep{kennicutt98a} yields a very flat SED in the UV region used in their relation \citep[1500--2800\AA][]{kennicutt98a}. This relation gives $SFR(M_{\Sun}\ $yr$^{-1})=1.4\times10^{-28}L_{\nu}($ergs$\ $s$^{-1}\ $Hz$^{-1})$. We choose to estimate $L_{\nu}$ at the mid-point of their UV spectrum (2150\AA) from our template fits, which is then used to estimate the unobscured SFR. The SFRs derived from our template fits are given in Table~\ref{tab:spectab}, where uncertainties in the UV SFRs are dominated by uncertainties in the template fits while those in the IR SFRs are dominated by the scatter in the scaling relations.  \citet{kennicutt98a} note that this UV--SFR relation has the benefit of directly 
tracking emission from young stellar atmospheres, but would also be quite sensitive to both extinction and variation 
of the IMF.  The \citet{kennicutt98} relation between IR luminosity and the SFR likewise assumes a Salpeter IMF 
with continuous star formation, but it also assumes that all of the bolometric luminosity is reprocessed in the 
infrared.

\begin{figure}
  \begin{center}
    \plotone{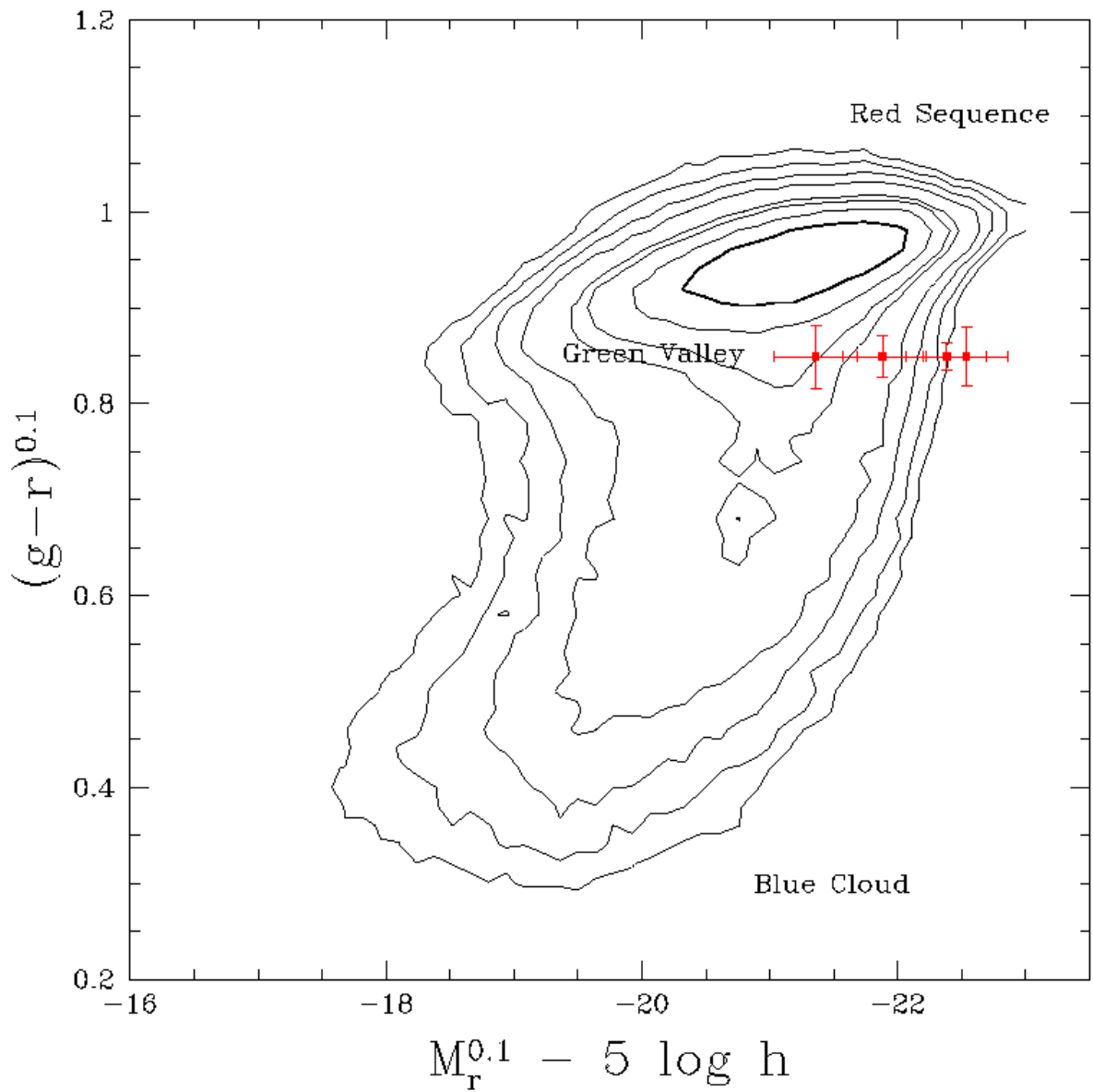}
    \caption{Color-magnitude diagram of SDSS galaxies from the SDSS Data Release 6 \citep{adelman08} with the points for 
the four images of SDSS~J1004+4112 in red.  The systematic effects on the luminosity from estimates of the magnification 
for each lensed image is clearly seen from the similarity in color as compared to the factor of $\simeq2$ differences in the 
estimates of the magnification-corrected luminosity.}
    \label{fg:cmd}
   \end{center}
\end{figure}

We combined the template models with the results of \citet{bell03} to estimate the stellar mass of the 
host galaxy.  \citet{bell03} assumed a universal ``diet Salpeter'' Initial Mass Function (IMF) and a variety of star 
formation histories to simulate SED templates, which they then fit to a large sample of SDSS galaxies 
to estimate mass-to-light ratios as a function of rest frame colors. They took these M/L ratios, in 
combination with the measured colors of the galaxies, to derive relationships between colors and mass-to-light ratios, 
$\log (M/L) = a_{\lambda}+(b_{\lambda}\times $color$)$, as detailed in \citet{bell03} (Table 7). We 
assume a Kroupa IMF, which better represents a normal stellar population, and this introduces a $-0.15$ 
dex correction to the value of $a_{\lambda}$. We estimate the rest frame $(g-r)$ color and K-band luminosity from our template fits, and then 
use the \citet{bell03} K-band parameters ($a_K = -0.359$ and $b_K = 0.197$) to estimate the mass-to-light ratio.  
This leads to an estimated host K-band 
$\log(M/L)=(-0.19\pm0.28)$(M/L)$_{\sun}$, with uncertainties dominated by random scatter in the \citet{bell03} calibration and in the color derived from the template fits.  The estimated rest frame $(g-r)\simeq 0.85 $ 
color of the host puts it in the ``green valley''between the star forming ``blue cloud'' and the 
``red sequence'' \citep[see e.g. ][]{strateva01,blanton03}, as seen in Figure~\ref{fg:cmd}.

The estimated MgII and C[IV] line FWHM are 49 and 21 \AA, respectively \citep[][Morgan, private communication]{fohlmeister08}.
These both indicate a black hole mass of $\log(M_{BH}/\mathrm{M}_{\sun})\simeq 8.6$ based on the scalings of 
\citet{mcclure02} for MgII and \citet{vestergaard06} for C[IV].  We have also applied the revised normalization 
of \citet{onken04} to the MgII estimate.  The estimated magnification-corrected luminosity at rest-frame at 
$1350$\AA\ is $2.0\times10^{45}$~erg/s based on power-law fits to the B, V, and I HST images.  The uncertainties in the $M_{BH}$ 
estimate are dominated by systematics, principally the ~0.3 dex uncertainty typical of $M_{BH}$ estimates from line 
widths and ~0.15 dex from the magnification uncertainties.  Nonetheless, the excellent agreement between the MgII and C[IV] mass 
estimates ($\log(M_{BH}/\mathrm{M}_{\sun})\ =$ 8.62 and 8.56, respectively) is reassuring.
If we estimate a black hole mass from the rest frame V-band host luminosity of 
$(2.07\pm0.03)\times10^{43}$~ergs/s, using the relation of \citet{gultekin09}, 
we find a black hole mass of $\log(M_{BH}/M_{\sun})=8.74\pm0.21$ that agrees 
well with the estimates from the line widths.  Note, however, that we have no 
estimate of the fraction of the galactic luminosity that comes from the host's bulge, 
as used in the \citet{gultekin09} relation, since decomposing the host galaxy into bulge 
and disk components is made impossible by the geometry of the lensing of the host and the 
high luminosity of the quasar.  The Eddington luminosity for such a black hole is
\begin{equation}\label{eq:ledd}
L_{Edd} = 5.7 \times 10^{12}  \left(\frac{M_{BH}}{10^{8.6} \mathrm{M}_{\sun}}\right)\ \mathrm{L}_{\sun}.
\end{equation}
From our template models we can estimate the 0.1--24$\mu$m luminosity of the black hole (see Table \ref{tab:spectab}).  We use 
the 3$\mu$m to $L_{IR}$ analysis from \S 2.6 of \citet{gallagher07}, applied to the Bo\"otes field 
AGNs to estimate a bolometric correction of BC$\simeq$1 between this luminosity and the bolometric 
luminosity \citep[for an in depth discussion, see][]{assef09}.

Based on these scalings, a weighted average over the different lensed images, and assuming a Kroupa IMF, we estimate that the total (obscured plus unobscured) star formation 
rate is $\log(\dot{M}/$M$_{\odot}$~yr$^{-1})=1.21\pm0.26$ compared to a stellar mass of 
$\log (M_{\star}/M_{\sun})=11.09\pm0.28$.  The uncertainties in these quantities are dominated by the scatter in 
the IR scaling relations.  Aside from the systematic and random uncertainties in the scalings used to determine 
the SFR \citep[40\%,][]{kennicutt98,bavouzet08} and stellar mass \citep[26\%,][]{bell03}, the biggest uncertainties 
arise from the magnification estimates.  The IRAC quasar flux ratios are probably a reasonable estimate of the 
intrinsic flux ratios because we expect (and observe) little variability at these wavelengths, as quasar variability diminishes to longer wavelengths \citep{vandenberk04}, extinction effects will be negligible, and the mid-IR emission region is too large to be strongly microlensed \citep[e.g.][on microlensing in HE~1104-1805]{poindexter07}.  These IR flux ratios are B/A$\sim$0.76, C/A$\sim$0.63, and D/A$\sim$0.32 compared to 0.67, 0.34, and 0.27 
respectively from the \citet{inada08} models (Oguri, private communication). Much of this will be incorporated into 
the uncertainties estimated from the scatter between the various lensed images and masks.  The host SED has roughly 
equal contributions from the E and Sbc templates and little contribution from the Irr template, independent of the 
image or region fit.  While dust could obscure the optical/UV emission of young stars in the Irr template (see Figure 
\ref{fg:sed_joint}), we would not find a good fit to the host using an obscured Irr template. Note that the inner 
(QSO mask) and outer (Host mask) regions of the host galaxy seem to contain a similar number of stars and have similar 
specific star formation rates.

We can also compare these inferences about the stars to those for the black hole.  The black hole represents a 
mass fraction of $\log(M_{BH}/M_{\star}) = -2.49 \pm 0.28 $ compared to the stars, which is marginally inconsistent with local 
estimates of $\log(M_{BH}/M_{\star}) = -2.85 \pm 0.12$ \citep{haring04}.  Our result is in better agreement with the Peng et al. (2006a) estimate that the $M_{BH}/M_*$ relation is $4_{-1}^{+2}$ ($\simeq 0.6$~dex) 
larger at $z=1.7$ than locally (i.e. $\log(M_{BH}/M_{\star})\simeq -2.25$). Assuming that the evolution in this relationship found by \citet{peng06a} is correct, the agreement of the black hole masses estimated from the line widths and the host luminosity is then a coincidence in which the effect of evolution in the relation is balanced by our over-estimation of the bulge luminosity. Finally, we note that after including our estimate of 
the bolometric correction, we find that $L_{BH}/L_{Edd} \simeq  0.24\pm0.05$, so the black hole is radiating at a 
significant fraction of its Eddington limit, as is typical of quasars at this epoch \citep[e.g.][]{kollmeier06}.  
The quasar may be moderately extincted, as we find best fits where the quasar template is reddened by 
$E(B-V)\simeq0.1$, 0.1, 0.15, and 0.0 magnitudes for the A-D images, respectively.

In summary, both the host galaxy and quasar in SDSS~J1004+4112 have relatively unremarkable properties.  The one 
exception is that the host galaxy lies in the ``green valley.'' It is unclear from our template fits whether this 
galaxy is then in transition from being a star forming galaxy in the ``blue cloud'' to an old, red, and dead galaxy 
on the red sequence, or a red sequence galaxy with a recent burst of star formation that moved its color blue-ward. 
This galaxy's location in the CMD is consistent with the observation of \citet{hickox09} that many X-ray AGN with the X-ray luminosity of SDSS~J1004+4112 
\citep[$\simeq2\times10^{43}$~ergs/s,][]{ota06,lamer06} lie in the green valley, while radio AGN tend to lie on 
the red sequence and mid-IR selected AGN tend to lie in the blue cloud. The extreme extension of the host galaxy 
should also make it possible to obtain spectroscopic observations of the host galaxy, potentially allowing measurement 
of the dynamical mass or metallicity.

\acknowledgments

We would like to thank C.\ Morgan and C.\ Peng for their comments and help on the estimated 
black hole mass.  We would also like to thank D.\ Maoz and M.\ Oguri for their comments.
This work is based in part on observations made with the NASA/ESA {\it Hubble Space Telescope}.  
Support for programs GO-9744, 10509 and 10716 was provided by NASA through a grant from the Space 
Telescope Science Institute, which is operated by AURA, Inc., under NASA contract NAS5-2655.  It is 
also based in part on observations made with the Spitzer Space Telescope, AO-20451, which is operated by the 
Jet Propulsion Laboratory, California Institute of Technology under a contract with NASA.  Support 
for this program SST-20277 was provided by NASA through an award issued by JPL/Caltech.  CSK is also 
supported by NSF grant AST-0708082.

\begin{deluxetable}{r c c c c c}
\tablecolumns{6}
\tablewidth{0pt}
\tablecaption{Observations of SDSS J1004+4112 \label{tab:obs}}
\tablehead{
\colhead{Program} & \colhead{Instrument} & \colhead{Filter} & \colhead{Date} & \colhead{Exposure Time (sec)} & \colhead{Depth}
}
\startdata
HST-9744 &  NICMOS/NIC2 & H(F160W) & 2004/04/28 & \ \ 2688* & 21.77\,\,\\
\phn & NICMOS/NIC2 & H(F160W) & 2004/10/09 & \ 2688 & 21.70\,\,\\
HST-10716 & NICMOS/NIC2 & H(F160W) & 2006/10/22 & \ 2688 & 21.45\,\,\\
HST-10509 & ACS/WFC & B(F435W) & 2005/12/13 & 13378 & 26.26\,\,\\ 
HST-9744 & ACS/WFC & V(F555W) & 2004/01/28 & \ 2025  & 25.22\,\,\\
HST-10509 & ACS/WFC & V(F555W) & 2005/12/12 & \ 7978  & 26.30\,\,\\
HST-9744 & ACS/WFC & I(F814W) & 2004/04/28 & \ 2025  & 24.29\,\,\\ 
HST-10509 & ACS/WFC & I(F814W) & 2005/12/12 & \ 5360  & 24.41\,\,\\
SST-20277 & IRAC & 3.6--8.0$\mu$m & 2005/12/08, & \ 3485  & 20.28$^1$\\
\phn & \phn & \phn & 2005/12/26, &   & 19.96$^2$\\
\phn & \phn & \phn & 2006/11/25 &  & 18.37$^3$\\
\phn & \phn & \phn & \phn &  & 17.76$^4$\\
\enddata

\tablecomments{*This observation includes only images A and B.
Depths are 5-$\sigma$ rms noise, in Vega magnitudes, in a $\sim$1~arcsec$^2$ aperture for the 
HST bands and $\sim$4~arcsec$^2$ aperture for the SST/IRAC images.
$^1$ IRAC Channel 1, $^2$ IRAC Channel 2, $^3$ IRAC Channel 3, $^4$ IRAC 
Channel 4.}
\end{deluxetable}

\begin{deluxetable}{c c c c c c c c c c c c}

\rotate

\tablecolumns{12}
\tablewidth{0pt}
\tablecaption{SDSS J1004+4112 Magnitudes\label{tab:fluxtab}}
\tabletypesize{\tiny}
\tablehead{
\colhead{} & \colhead{} & \multicolumn{3}{c}{Image A} & \multicolumn{3}{c}{Image B} & \multicolumn{3}{c}{Image C} & \colhead{Image D}  \\
\colhead{Filter} & \colhead{Date} & \colhead{host} & \colhead{qso} & \colhead{joint} & \colhead{host} & \colhead{qso} & \colhead{joint} & \colhead{host} & \colhead{qso} & \colhead{joint} & \colhead{joint}
}

\startdata
B & 2005/12/13 & 28.18$\pm$0.81 & 23.44$\pm$0.03 & 23.42$\pm$0.01 & 28.55$\pm$0.67 & 23.24$\pm$0.03 & 23.22$\pm$0.01 & 26.43$\pm$0.86 & 22.67$\pm$0.57 & 22.64$\pm$0.45 & 22.87$\pm$0.36 \\
V & 2004/01/28 & 27.93$\pm$1.12 & 23.36$\pm$0.15 & 23.34$\pm$0.13 & 27.39$\pm$1.18 & 23.29$\pm$0.15 & 23.26$\pm$0.13 & 26.34$\pm$1.25 & 22.65$\pm$0.15 & 22.61$\pm$0.15 & 22.41$\pm$0.14 \\
V & 2005/12/12 & 27.76$\pm$1.11 & 23.29$\pm$0.10 & 23.27$\pm$0.08 & 27.59$\pm$1.19 & 23.27$\pm$0.11 & 23.24$\pm$0.08 & 26.69$\pm$1.09 & 22.62$\pm$0.11 & 22.59$\pm$0.08 & 22.73$\pm$0.08 \\
I & 2004/04/28 & 25.52$\pm$0.97 & 22.30$\pm$0.16 & 22.26$\pm$0.16 & 25.39$\pm$0.81 & 22.16$\pm$0.16 & 22.11$\pm$0.15 & 24.73$\pm$0.65 & 21.48$\pm$0.16 & 21.41$\pm$0.16 & 21.86$\pm$0.16 \\
I & 2005/12/12 & 25.78$\pm$0.73 & 22.42$\pm$0.10 & 22.36$\pm$0.04 & 25.57$\pm$0.72 & 22.27$\pm$0.12 & 22.21$\pm$0.04 & 24.65$\pm$0.61 & 21.83$\pm$0.24 & 21.76$\pm$0.19 & 21.64$\pm$0.03 \\
H & 2004/04/28 & 22.43$\pm$0.23 & 21.26$\pm$0.10 & 21.02$\pm$0.11 & 22.36$\pm$0.14 & 21.09$\pm$0.11 & 20.86$\pm$0.11 & \nodata & \nodata & \nodata & \nodata \\
H & 2004/10/09 & 22.41$\pm$0.41 & 20.93$\pm$0.08 & 20.66$\pm$0.13 & 21.89$\pm$0.40 & 20.88$\pm$0.09 & 20.47$\pm$0.14 & 21.36$\pm$0.40 & 20.05$\pm$0.10 & 19.64$\pm$0.15 & 19.85$\pm$0.13 \\
H & 2006/10/22 & 22.38$\pm$0.15 & 21.16$\pm$0.10 & 20.87$\pm$0.09 & 21.94$\pm$0.33 & 20.86$\pm$0.10 & 20.48$\pm$0.13 & 21.79$\pm$0.37 & 20.53$\pm$0.08 & 20.21$\pm$0.11 & 20.20$\pm$0.10 \\
3.6$\mu$m & 2005/12/08 & 19.94$\pm$0.33 & 18.69$\pm$0.01 & 18.30$\pm$0.05 & 19.42$\pm$0.21 & 18.60$\pm$0.03 & 18.07$\pm$0.09 & 19.05$\pm$0.58 & 18.11$\pm$0.04 & 17.64$\pm$0.16 & 17.79$\pm$0.17 \\
3.6$\mu$m & 2005/12/26 & 19.80$\pm$0.17 & 18.72$\pm$0.02 & 18.29$\pm$0.07 & 19.41$\pm$0.33 & 18.65$\pm$0.02 & 18.10$\pm$0.10 & 19.18$\pm$0.34 & 18.05$\pm$0.02 & 17.62$\pm$0.06 & 17.77$\pm$0.04 \\
3.6$\mu$m & 2006/11/25 & 19.80$\pm$0.22 & 18.86$\pm$0.06 & 18.42$\pm$0.10 & 19.37$\pm$0.32 & 18.67$\pm$0.06 & 18.10$\pm$0.07 & 18.96$\pm$0.22 & 18.07$\pm$0.06 & 17.58$\pm$0.08 & 17.71$\pm$0.06 \\
4.5$\mu$m & 2005/12/08 & 19.12$\pm$0.23 & 17.87$\pm$0.01 & 17.46$\pm$0.06 & 18.90$\pm$0.11 & 17.74$\pm$0.02 & 17.32$\pm$0.03 & 18.14$\pm$0.07 & 17.36$\pm$0.02 & 16.83$\pm$0.03 & 17.07$\pm$0.04 \\
4.5$\mu$m & 2005/12/26 & 19.01$\pm$0.22 & 17.92$\pm$0.02 & 17.47$\pm$0.03 & 18.86$\pm$0.08 & 17.81$\pm$0.02 & 17.36$\pm$0.03 & 18.16$\pm$0.15 & 17.34$\pm$0.02 & 16.81$\pm$0.07 & 17.08$\pm$0.02 \\
4.5$\mu$m & 2006/11/25 & 19.04$\pm$0.22 & 18.01$\pm$0.05 & 17.59$\pm$0.05 & 18.89$\pm$0.04 & 17.85$\pm$0.05 & 17.43$\pm$0.05 & 18.07$\pm$0.21 & 17.41$\pm$0.05 & 16.87$\pm$0.06 & 16.99$\pm$0.05 \\
5.8$\mu$m & 2005/12/08 & 18.92$\pm$0.32 & 17.03$\pm$0.01 & 16.71$\pm$0.03 & 18.60$\pm$0.15 & 16.96$\pm$0.02 & 16.62$\pm$0.06 & 18.09$\pm$0.23 & 16.49$\pm$0.01 & 16.15$\pm$0.02 & 16.40$\pm$0.03 \\
5.8$\mu$m & 2005/12/26 & 19.08$\pm$0.40 & 17.06$\pm$0.01 & 16.76$\pm$0.04 & 18.87$\pm$0.18 & 16.96$\pm$0.01 & 16.65$\pm$0.03 & 18.15$\pm$0.13 & 16.46$\pm$0.01 & 16.12$\pm$0.02 & 16.41$\pm$0.02 \\
5.8$\mu$m & 2006/11/25 & 18.95$\pm$0.40 & 17.12$\pm$0.04 & 16.90$\pm$0.06 & 18.77$\pm$0.22 & 16.91$\pm$0.04 & 16.70$\pm$0.05 & 18.04$\pm$0.15 & 16.42$\pm$0.04 & 16.22$\pm$0.07 & 16.35$\pm$0.04 \\
8.0$\mu$m & 2005/12/08 & 18.53$\pm$0.17 & 15.93$\pm$0.01 & 15.89$\pm$0.01 & 18.27$\pm$0.48 & 15.85$\pm$0.01 & 15.80$\pm$0.02 & 17.96$\pm$0.32 & 15.31$\pm$0.01 & 15.38$\pm$0.01 & 15.73$\pm$0.02 \\
8.0$\mu$m & 2005/12/26 & 18.51$\pm$0.18 & 16.07$\pm$0.01 & 15.88$\pm$0.02 & 18.45$\pm$0.30 & 16.00$\pm$0.01 & 15.85$\pm$0.03 & 17.87$\pm$0.14 & 15.49$\pm$0.01 & 15.38$\pm$0.01 & 15.71$\pm$0.02 \\
8.0$\mu$m & 2006/11/25 & 18.46$\pm$0.08 & 15.87$\pm$0.03 & 16.02$\pm$0.03 & 18.27$\pm$0.23 & 15.73$\pm$0.03 & 15.91$\pm$0.04 & 17.78$\pm$0.09 & 15.11$\pm$0.03 & 15.45$\pm$0.03 & 15.66$\pm$0.04 \\

\enddata

\tablecomments{These are Vega magnitudes corrected for magnification by the lens based on the models of \citet[][Oguri private communication]{inada08}. The fluxes in the QSO mask contain a portion of the host galaxy. The QSO and joint fluxes are adjusted to account for time-dependent variability in the quasar using corrections from the monitoring data of \citet{fohlmeister08}.  The 2004/04/28 NICMOS H-band observation included only images A and B.}

\end{deluxetable}
\begin{deluxetable}{c c c c c c c c c c}

\rotate

\tablecolumns{10}
\tablewidth{0pt}
\tablecaption{Magnification Corrected Properties \label{tab:spectab}}
\tablehead{
\colhead{Image} & \colhead{Mask} & \colhead{E} & \colhead{Sbc} & \colhead{Irr} &
\colhead{QSO} & \colhead{SFR$_{IR}$} & \colhead{SFR$_{UV}$} & $(M/L)_K$ & \colhead{M$_{\star}$} \\
\cline{3-6} \\
\colhead{} & \colhead{} & \multicolumn{4}{c}{(Luminosity in units of $10^{10} $L$_{\Sun}$)} & 
\colhead{(M$_{\Sun}\,$yr$^{-1}$)} & \colhead{(M$_{\Sun}\,$yr$^{-1}$)} & Solar Units & \colhead{$(10^{10} $M$_{\Sun})$}
}

\startdata
A   & host   & 1.05$\pm$0.64 & \phantom{1}2.84$\pm$0.95 & 0.038$\pm$0.120 & $\equiv$0 & \phantom{1}8.8$\pm$4.0\phantom{1} & $0.36\pm0.14$ &  0.605$\pm$0.100
& \phantom{1}4.3$\pm$0.7 \\
    & qso    & 1.35$\pm$0.01 & \phantom{1}1.93$\pm$0.01 & 0.104$\pm$0.001 & 145.6$\pm$0.2 & \phantom{1}6.5$\pm$2.4\phantom{1} & $0.19\pm0.01$ & 0.597$\pm$0.105
& \phantom{1}1.9$\pm$0.4 \\ 
    & joint  & 2.87$\pm$0.06 & \phantom{1}4.09$\pm$0.09 & 0.219$\pm$0.005 & 125.7$\pm$0.7 & 12.1$\pm$4.5\phantom{1} & $0.80\pm0.02$ & 0.578$\pm$0.100
& \phantom{1}8.4$\pm$1.0 \\ 
B   & host   & 2.02$\pm$1.04 & \phantom{1}1.86$\pm$1.72 & 0.049$\pm$0.133 & $\equiv$0 & \phantom{1}6.3$\pm$5.1\phantom{1} & $0.29\pm0.11$ & 0.612$\pm$0.100
& \phantom{1}4.9$\pm$0.6 \\ 
    & qso    & 1.40$\pm$0.01 & \phantom{1}1.99$\pm$0.01 & 0.107$\pm$0.001 & 167.0$\pm$0.2 & \phantom{1}6.7$\pm$2.5\phantom{1} & $0.53\pm0.02$ & 0.591$\pm$0.101
& \phantom{1}5.6$\pm$0.8 \\ 
    & joint  & 4.69$\pm$0.06 & \phantom{1}6.67$\pm$0.09 & 0.358$\pm$0.005 & 108.9$\pm$0.8 & 18.2$\pm$6.8\phantom{1} & $1.30\pm0.02$ & 0.575$\pm$0.100 
& 13.6$\pm$1.5 \\ 
C   & host   & 4.06$\pm$1.49 & \phantom{1}2.57$\pm$2.25 & 0.685$\pm$0.448 & $\equiv$0 & \phantom{1}8.9$\pm$6.4\phantom{1} & $1.22\pm0.54$ & 0.610$\pm$0.100
& \phantom{1}9.0$\pm$1.4 \\ 
    & qso    & 3.00$\pm$0.02 & \phantom{1}4.26$\pm$0.02 & 0.229$\pm$0.010 & 250.7$\pm$0.6 & 12.6$\pm$4.7\phantom{1} & $0.70\pm0.01$ & 0.595$\pm$0.103
& \phantom{1}7.3$\pm$0.8 \\ 
    & joint  & 8.57$\pm$0.17 & 12.19$\pm$0.24 & 0.655$\pm$0.013 & 166.6$\pm$1.7 & 30.1$\pm$11.1 & $2.38\pm0.05$ & 0.590$\pm$0.101
& 24.9$\pm$2.8 \\ 
D   & joint  & 7.42$\pm$0.07 & 10.55$\pm$0.10 & 0.567$\pm$0.005 & 126.2$\pm$1.4 & 26.7$\pm$9.9\phantom{1} & $2.06\pm0.02$ & 0.621$\pm$0.102
& 21.6$\pm$2.2 \\ 
Avg & host   & 1.64$\pm$0.51 & \phantom{1}2.60$\pm$0.78 & 0.067$\pm$0.087 & $\equiv$0 & 8.1$\pm$2.8 & $0.35\pm0.52$ & 0.609$\pm$0.058
& \phantom{1}5.1$\pm$0.5 \\ 
    & qso    & 1.81$\pm$0.03 & \phantom{1}1.96$\pm$0.03 & 0.165$\pm$0.003 & 155.0$\pm$0.6 & 4.8$\pm$1.0 & $0.44\pm0.26$ & 0.594$\pm$0.059
& \phantom{1}3.4$\pm$0.3 \\ 
    & joint  & 5.54$\pm$0.04 & \phantom{1}6.01$\pm$0.04 & 0.505$\pm$0.003 & 124.0$\pm$0.5 & 15.0$\pm$3.0\phantom{1} & $1.39\pm0.72$ & 0.591$\pm$0.050
& 12.3$\pm$0.7 \\ 

\enddata

\tablecomments{Contribution of each template to the SED. These are corrected for magnification based on the models of \citet[][Oguri private communication]{inada08}. The star formation rates (SFR) and stellar mass are 
estimated as described in \S \ref{sec:analysis}. Note that the average values at the bottom of the table (Avg) are uncertainty weighted averages of the corresponding values for the individual components.}
\end{deluxetable}

\begin{thebibliography}{100}

\bibitem[Adelman-McCarthy et al., 2008]{adelman08}
Adelman-McCarthy, J.K., et al.\ 2008, ApJS, 175, 297

\bibitem[Assef et al., 2008]{assef08}
Assef, R.J., et al.\ 2008, \apj, 676, 286

\bibitem[Assef et al., 2009]{assef09}
Assef, R.J., et al.\ 2009, in preparation

\bibitem[Bavouzet et al., 2008]{bavouzet08}
Bavouzet, N., et al.\ 2008, arXiv:0712.0965v1 [astro-ph]

\bibitem[Bell et al., 2003]{bell03}
Bell, E.F., et al.\ 2003, \apj, 149, 289

\bibitem[Blanton et al., 2003]{blanton03}
Blanton, M.R., et al.\ 2003, \apj, 594, 186B

\bibitem[Caputi et al., 2007]{caputi07}
Caputi, K.I., et al.\ 2007, \apj, 660, 97C

\bibitem[Di Matteo et al., 2008]{dimatteo08}
Di Matteo, T., et al.\ 2008, \apj, 676, 33

\bibitem[Fohlmeister et al., 2007]{fohlmeister07}
Fohlmeister, J., et al.,\ 2007, \apj, 662, 62

\bibitem[Fohlmeister et al., 2008]{fohlmeister08}
Fohlmeister, J., et al.\ 2008, \apj, 676, 761

\bibitem[Fruchter et al., 2002]{fruchter02}
Fruchter et al.\ 2002, PASP, 114, 144

\bibitem[Gallagher et al., 2007]{gallagher07}
Gallagher, S.C., et al.\ 2007, \apj, 661, 30-37

\bibitem[G\"ultekin et al., 2009]{gultekin09}
G\"ultekin, K., et al.\ 2009, arXiv:0903.4897

\bibitem[H\"aring \& Rix, 2004]{haring04}
H\"aring, N., Rix, H.-W.\ 2004, \apj, 604, 89-92

\bibitem[Hickox et al., 2009]{hickox09}
Hickox, R.C., et al.\ 2009, arxiv: 0901.4121

\bibitem[Hopkins et al., 2005a]{hopkins05a}
Hopkins, P., et al.\ 2005a, \apj, 625, 71

\bibitem[Hopkins et al., 2005b]{hopkins05}
Hopkins, P., et al.\ 2005b, \apj, 630, 705

\bibitem[{Hopkins, Narayan, \& Hernquist}, 2006]{hopkins06}
Hopkins, P., Narayan, R.,  Hernquist, L.\ 2006, \apj, 643, 641

\bibitem[Hopkins et al., 2006]{hopkins06b}
Hopkins, P., et al.\ 2006, ApJS, 163, 1

\bibitem[Hopkins et al., 2008]{hopkins08}
Hopkins, P., et al.\ 2008, ApJS, 175, 356

\bibitem[Inada et al., 2003]{inada03}
Inada, N., et al.\ 2003, Nature, 426, 810-812

\bibitem[Inada et al., 2005]{inada05}
Inada, N., et al., \ 2005, PASJ, 57, 71

\bibitem[Inada et al., 2008]{inada08}
Inada, N., et al.\ 2008, PASJ, 60, 27

\bibitem[Jannuzi \& Dey, 1999]{januzzi99}
Jannuzi,  B.T., Dey, A.\ 1999 AAS, 195, 1207J

\bibitem[Kauffmann \& Heckman, 2005]{kauffmann05}
Kauffmann, G., Heckman, T.\ 2005, RSPTA, 363, 621

\bibitem[Kennicutt, 1998a]{kennicutt98a}
Kennicutt Jr., R.C.\ 1998a, ARA\&A, 36, 189K

\bibitem[Kennicutt, 1998b]{kennicutt98}
Kennicutt Jr., R.C.\ 1998b, \apj, 498, 541

\bibitem[Kochanek et al., in prep.]{kochanek08}
Kochanek, C.S., et al.\ 2008, in preparation

\bibitem[Kollmeier et al., 2006]{kollmeier06}
Kollmeier, J., et al.\ 2006, \apj, 648, 128

\bibitem[Lamer et al., 2006]{lamer06}
Lamer et al.\ 2006, A\&A, 454, 493

\bibitem[Lauer et al., 2007]{lauer07}
Lauer, T., et al.\ 2007, \apj, 670, 249

\bibitem[Leh\'ar et al., 2000]{lehar00}
Leh\'ar, J., et al.\ 2000, \apj, 536, 584

\bibitem[Madau et al., 1998]{madau98}
Madau, P., et al.\ 1998, \apj, 498, 106

\bibitem[Marconi \& Hunt, 2003]{marconi03}
Marconi, A., Hunt, L.\ 2003, \apj, 589, 21

\bibitem[McClure \& Jarvis, 2002]{mcclure02}
McClure \& Jarvis, 2002, MNRAS, 337, 109

\bibitem[Oguri, private communication]{oguripc}
Oguri, M., Private communication with C.S. Kochanek 2008/07/22

\bibitem[Onken et al., 2004]{onken04}
Onken, C., et al.\ 2004, \apj, 615, 645

\bibitem[Ota et al., 2006]{ota06}
Ota, N., et al.,\ 2006, \apj, 647, 215

\bibitem[Peng et al., 2006a]{peng06a}
Peng, C.Y., et al.\ 2006a, \apj, 640, 114

\bibitem[Peng et al., 2006b]{peng06b}
Peng, C.Y., et al.\ 2006b, \apj, 649, 616

\bibitem[Peng et al., 2006c]{peng06c}
Peng, C.Y., et al.\ 2006c, New Astron.Rev. 50, 689-693

\bibitem[Poindexter et al., 2007]{poindexter07}
Poindexter, S., et al.\ 2007, \apj, 660, 146

\bibitem[Sharon et al., 2005]{sharon05}
Sharon, K., et al.\ 2005, \apj, 629, 73

\bibitem[Sijacki et al., 2007]{sijacki07}
Sijacki, D., et al.\ 2007, MNRAS, 380, 877-900

\bibitem[Strateva et al., 2001]{strateva01}
Strateva, I., et al.\ 2001, AJ, 122, 1861S

\bibitem[Vanden Berk et al., 2004]{vandenberk04}
Vanden Berk et al.\ 2004, \apj, 601, 692

\bibitem[Vestergaard \& Peterson, 2006]{vestergaard06}
Vestergaard \& Peterson, 2006, \apj, 641, 689

\end{thebibliography}
\end{document}